\documentclass{emulateapj}

\newcommand{\xte}{{\it RXTE}}
\newcommand{\chandra}{{\it Chandra}}
\newcommand{\sax}{{\it BeppoSAX}}
\newcommand{\eps}{{\rm ergs\,s^{-1}}}
\newcommand{\epcs}{{\rm ergs\,cm^{-2}\,s^{-1}}}
\newcommand{\epc}{{\rm ergs\,cm^{-2}}}

\newcommand{\fper}{$F_p$}

\newcommand{\srcb}{SAX~J1808.4$-$3658}

\newcommand{\srcd}{XTE~J1751$-$305}

\newcommand{\srcf}{IGR~J00291+5934}

\newcommand{\catlabel}{G06}	

\slugcomment{Accepted by ApJ}	

\shorttitle{Helium-rich bursts from SAX~J1808.4$-$3658}
\shortauthors{Galloway and Cumming}

\begin{document}

\title{Helium-rich thermonuclear bursts and 
  the distance to the accretion-powered millisecond pulsar
SAX~J1808.4$-$3658}

\author{Duncan K. Galloway\altaffilmark{1,2} and
   Andrew Cumming\altaffilmark{3}} 
\affil{Kavli Institute for Astrophysics and Space Research,
  Massachusetts Institute of Technology, Cambridge, MA 02139}

\altaffiltext{1}{present address: School of Physics, University of Melbourne,
  Victoria, Australia; email: D.Galloway@physics.unimelb.edu.au}
\altaffiltext{2}{Centenary Fellow}
\altaffiltext{3}{Physics Department, McGill University,
  Montreal, QC, H3A 2T8, Canada; email: cumming@physics.mcgill.ca}

\begin{abstract}
We analysed {\it Rossi X-ray Timing Explorer}\/ observations of the
accretion-powered 401~Hz pulsar \srcb, in order to precisely determine the
source distance. While the fluences for the five transient outbursts observed from
1996 were constant to within the uncertainties, the outburst interval
varied signficantly, so that the time-averaged flux (and accretion rate)
decreased by around 40\%.
By equating the time-averaged X-ray flux with the expected mass transfer
rate from gravitational radiation, we derived a lower limit on the
distance of 3.4~kpc. Combined with an upper limit from assuming that the
four radius-expansion thermonuclear bursts observed during the 2002
October outburst reached at most the Eddington limit for a pure He
atmosphere, we found that the probable distance range for the source is
3.4--3.6~kpc.
The implied inclination, based on the optical/IR properties of the
counterpart, is $i\la30^\circ$.

We compared the properties of the bursts with an ignition model. The time
between bursts was long enough for hot CNO burning to significantly deplete
the accreted hydrogen, so that ignition occurred in a pure helium layer
underlying a stable hydrogen burning shell. This is the first time that
this burning regime has been securely observationally identified. The
observed energetics of the bursts give a mean hydrogen fraction at
ignition of $\left<X\right>\approx0.1$, and require that the accreted
hydrogen fraction $X_0$ and the CNO metallicity $Z_{\rm CNO}$ are
related by $Z_{\rm CNO}\approx 0.03\ (X_0/0.7)^2$. We show that in
this burning regime, a measurement of the burst recurrence time and
energetics allows the local accretion rate onto the star to be determined
independently of the accreted composition, giving a new method for
estimating the source distance
which is in good agreement with our other estimates.
\end{abstract}


\section{Introduction}

The population of accretion-powered millisecond pulsars (MSPs) has grown
rapidly over the last three years, with the discovery of six new examples
to bring the total sample to seven. \srcb\ remains the best-studied of the
seven sources, mainly due to the fact that it exhibits the most frequent
outbursts, every 2.5 years or so. The source was first discovered as a
faint transient with \sax\ \cite[]{zand98c}. Orbital phase-connection
across different outbursts has given the most precise determination of the
orbital parameters \cite[]{papitto05}; the mass donor, which orbits the
neutron star every 2.01~hours, is probably a brown dwarf heated by X-ray
emission during quiescence \cite[]{bc01}. The review of \cite{wij03a}
includes details of the observations undertaken during each outburst.
Optical observations of the counterpart V4580~Sgr \cite[]{gil99} have led
to limits on the inclination of $\cos i=0.65^{+0.23}_{-0.33}$ (90\%
confidence), assuming a distance of 2.5~kpc \cite[]{wang01}. The most
recent outburst was in 2005 June \cite[]{mark05b,wij05a}. The neutron star
has been detected in quiescence by \sax\/ \cite[]{stella00} and {\it
XMM-Newton}\/ \cite[]{campana02}, and a
\chandra\/ observation in outburst revealed the possible presence of a
narrow O {\sc vii} absorption line in addition to neutral absorption
features attributable to the ISM \cite[]{gal05a}. 

In this paper, we present a comprehensive study of the properties of
thermonuclear (type I) X-ray bursts and the persistent X-ray emission from
the accretion-powered MSP \srcb. We show that the sequence of X-ray bursts
observed during the 2002 October outburst can be understood as a series of
thermonuclear flashes driven by unstable helium burning in a layer of pure
helium underlying a stable hydrogen burning shell. This regime of burning
is predicted by X-ray burst models when the time between bursts is longer
than $\sim 1$ day, sufficient for the hot CNO cycle to exhaust the accreted hydrogen and leading to ignition in a pure helium environment 
\citep{fhm81,bil98a}. As far as we are aware, however, this regime has not been securely identified before. Short duration ($\lesssim 10\ {\rm s}$) X-ray bursts are observed from Atoll sources at X-ray luminosities $L_X\gtrsim 10^{37}\ {\rm erg\ s^{-1}}$, and are often referred to as ``helium-rich'' because they do not show the long tails characteristic of hydrogen burning. However, these luminosities are several times larger than those where pure helium ignition bursts are expected, and in addition, the bursts are not well-described by the standard theory, for example having irregular recurrence times and energetics that indicate additional burning of fuel between bursts
(e.g.~\citealt{bil98a}).

One of the aims of this study is to compare different methods used to
constrain the distance to \srcb\ and other MSPs. Two principal methods
have been used to date. First, three of the seven sources have shown
thermonuclear
X-ray bursts, which  have a maximum peak flux believed to correspond to
the Eddington limit at the surface \cite[e.g.][]{lew93,kuul03a}. The
uncertainty in distances determined by this method arises from uncertainty
in the mean hydrogen fraction at the photosphere, and the neutron star
mass. The distance to \srcb\ was initally estimated at around 4~kpc based
on the peak flux of a thermonuclear burst \cite[]{zand98c}. 
The likely distance range was later revised to between 2.5 and 3.3~kpc;
the decrease from 4~kpc resulted from a more accurate analysis of an
additional burst not detected in the original data, while the range arises
from the possible variation of the Eddington limit betweeen an H-rich and
pure He atmosphere \cite[]{zand01}. 

Second, the distance to a transient binary can be constrained by equating
the time-averaged X-ray flux (estimated from the outburst fluence divided
by the time since the previous outburst) and the flux predicted from the
expected mass transfer rate due to gravitational radiation $\dot{M}_{\rm
GR}$, which depends on the orbital period and the response of the mass
donor to the loss of material \cite[the mass-radius relation; see for
example][]{vvdh95}. Because pulse timing gives the projected semimajor
axis $a_X\sin i$ ($i$ is the inclination of the orbit), only a lower limit
on the companion mass is possible, which gives a lower limit to the
accretion rate and the distance. An important issue is the long term
X-ray flux behavior of the MSPs. The mass transfer rates in many
non-pulsing low-mass X-ray binaries (LMXBs) vary on time scales comparable
to MSP outburst intervals, and it is possible that the same is true for
the MSPs.
Presently this
remains an open question because the MSPs have uniformly low accretion
rates, and are transient, so that for most sources only one outburst has
been observed.

\begin{deluxetable}{lccccccc}
\tablecaption{Outburst properties and calculated distance limits for \srcb \label{outbursts}}
\tablewidth{240pt}
\tablehead{
  & \colhead{Start} & \colhead{Interval} & 
& 
  & \colhead{Distance}
\\
  \colhead{Outburst} &
  \colhead{(MJD)} &
  \colhead{(yr)\tablenotemark{a}} &
 \colhead{Fluence\tablenotemark{b}} & 
  \colhead{$\left<F_X\right>$\tablenotemark{c}} &
  \colhead{limit (kpc)
  }
}
\startdata
Sep 1996 & 50333 & $>0.67$ & $7.7\pm0.6$ & $<36$ & 1.4 \\ 
Apr 1998 & 50911 & 1.58 & $5.2\pm0.5$ & 10 & 2.5 \\ 
Jan 2000 & $\approx51547$ & 1.74 & $5.4\pm1.7$ & 9.6 & 2.6 \\ 
Oct 2002 & 52559 & 2.8 & $6.2\pm0.4$ & 7.1 & 3.1 \\ 
June 2005 & 53522 & 2.6 & $4.9\pm0.6$\tablenotemark{d} & 5.9 & 3.4 \\ 
\enddata
\tablenotetext{a}{The epoch for the outburst prior to the first known is
assumed to be earlier than the first ASM measurements (1996
January 6, or MJD~50088).}
\tablenotetext{b}{Estimated bolometric fluence in units of $10^{-3}\ \epc$}
\tablenotetext{c}{Estimated mean bolometric flux in units of $10^{-11}\
\epcs$.}
\tablenotetext{d}{The
fluence for the June 2005 outburst 
was estimated primarily from the ASM observations, since no public PCA data
from the first 20~d of the outburst
were available}
\end{deluxetable}

In this paper we also show that models of the sequence of X-ray bursts
from \srcb\ give a third estimate of the accretion rate onto the star and
therefore distance to the source. Remarkably, this distance estimate is
almost independent of the composition of the accreted material, the main input to the X-ray burst models. The plan of the paper is as follows. We present the observations and data analysis in \S 2. In \S\ref{distance} we calculate the lower limits to the distance obtained by equating the long-term time-averaged X-ray flux with the predicted mass transfer rate driven by gravitational radiation, addressing the issue of long term variations in the measured time-averaged persistent flux for the five outbursts observed so far. In \S\ref{bursters} we compare the burst properties with theoretical ignition models. We first show that the burst energetics require a specific correlation between the hydrogen mass fraction $X_0$ and the mass fraction of CNO elements $Z_{\rm CNO}$ in the accreted material. We then use this result to show that the measured recurrence times and $\alpha$ values imply an accretion rate onto the star that is independent of $X_0$ or $Z_{\rm CNO}$, and therefore gives a robust distance estimate. Finally,  in section \S\ref{pre}, we estimate the distance by equating the peak burst fluxes and the expected Eddington fluxes reached during radius-expansion bursts. We conclude in \S 4.

\section{Observations and analysis}

We analysed archival observations of 
\srcb\
made with the {\it Rossi X-ray Timing Explorer}\/
(\xte).  The proportional counter array \cite[PCA;][]{xte96} aboard \xte\/
consists of five identical proportional counter units (PCUs) sensitive to
photons with energies between 2--60~keV, with a field of view of $1^\circ$
and a total collecting area of $\approx6000\ {\rm cm^2}$. In order to
characterise the pulsar spectra up to $\approx100$~keV, we also used data
from the High-Energy X-ray Timing Experiment \cite[HEXTE;][]{hexte98},
which consists of two clusters of four NaI(Tl)/CsI(Na) phoswich
scintillation counters also with a field of view of $1^\circ$, sensitive
to photons with energies between 15--250~keV and with a total collecting
area of $1600\ {\rm cm^2}$.

We analysed the data with {\sc lheasoft} version 5.3, released 2003
November 17.
We extracted PCA Standard-2 data over each observation (excluding any
bursts), separately for each PCU. We fitted absorbed blackbody plus
powerlaw models in the energy range 2.5--25~keV, integrating the model to
derive the persistent source flux \fper. For selected (preferably long)
observations we made absorbed Comptonization (``{\tt comptt}'' in {\sc
xspec}) model fits to combined PCA and HEXTE
data, plus a Gaussian describing Fe K$\alpha$ emission around 6.4~keV,
where necessary.  We then generated an idealised response with 200
logarithmically-spaced energy bins between 0.1--200~keV and integrated the
model flux over this range, using the ratio of the 0.1-200~keV to
2.5--25~keV flux as a correction $c_{\rm bol}$ in estimating the
bolometric flux from each observation of each source.
The X-ray colors for 
\srcb\ were relatively constant throughout each outburst, and so we
adopted a constant value of
$c_{\rm bol}=2.12$.	
We used measurements of the burst properties 
tabled in the
catalog of
Galloway et al. (2006, in preparation; hereafter \catlabel).
These include the start time (measured as the
time the burst flux first exceeds 25\% of the maximum), bolometric fluence
$E_b$,
and the ratio of the integrated persistent flux to the fluence $\alpha$.
The latter parameter is subject to systematic errors due to intermediate
bursts which may be missed during source occultations occuring every
90~min orbit.

We 
compared the observed properties of the bursts with the ignition calculations
of \cite{cb00}, to which we refer the reader for further details.
We assumed a $1.4\ M_\odot$ neutron star with radius
$R_{\rm NS}=10\ {\rm km}$, giving a surface gravity
$g=(GM_{\rm NS}/R_{\rm NS}^2)(1+z)=2.45\times
10^{14}\ {\rm cm\ s^{-2}}$ 
where
$z$ is the redshift at the NS surface. For our choice of mass and radius,
$1+z=(1-2GM_{\rm NS}/R_{\rm NS}c^2)^{-1/2}=1.31$ \cite[cf.~][]{cott02}.
For a given accretion rate $\dot{m}$ (parameterised as a fraction of the
Eddington accretion rate $\dot{m}_{\rm Edd}\equiv8.8\times10^{-4}\ {\rm
g\,cm^{-2}}$) we calculated the temperature profile of the
accumulating layer of hydrogen and helium, adjusting its thickness until
the condition for a thermal runaway is met at the base. The temperature is mostly set by
hydrogen burning via the hot CNO cycle, and therefore the CNO mass
fraction $Z_{\rm CNO}$, which we refer to as the metallicity. The model also
includes
a flux from the crust $Q_b$. 
To calculate the burst energy, we assume complete burning of the H/He
fuel layer, and that the accreted material covers the whole surface of
the star. The total energy is then
$4\pi R^2 y Q_{\rm nuc}/(1+z)$,
where $y$ is the ignition column depth, and $Q_{\rm nuc}$ is
the energy per gram from nuclear burning.
We write $Q_{\rm nuc}=1.6+4\left<X\right>$ MeV per
nucleon, where $\left<X\right>$ is the mass-weighted mean hydrogen
fraction at ignition. This expression for $Q_{\rm nuc}$ assumes
$\approx 35$\% energy loss due to neutrinos during the rp process
\cite[e.g.][]{fuji87}, and gives $4.4$ MeV per nucleon for
approximately solar hydrogen abundance ($X=0.7$).

\begin{figure}
 \epsscale{1.2}
 \plotone{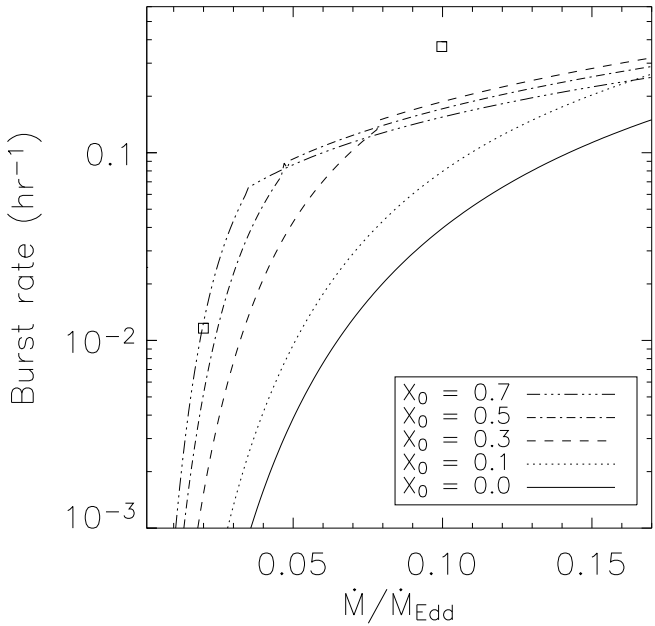}
 \figcaption[]{Predicted burst recurrence times according to the ignition
model for solar metallicity ($Z_{\rm CNO}=0.016$) in the accreted material.
The 
flux from the crust $Q_b$ 
is fixed at 0.3~MeV/nucleon, and we show curves for a range
of hydrogen fraction $X_0$ in the accreted material. The square symbols show
the prediction for the time-dependent model of \cite{woos03}.
 \label{model} }
\end{figure}

Figure \ref{model} shows the predicted burst rate as a function of
$\dot{m}$ for different accreted hydrogen mass fractions. The open squares
are the results of the time-dependent calculations of burst sequences by
\cite{woos03}. At low $\dot m\lesssim 0.05\ \dot m_{\rm Edd}$, 
the agreement between the two calculations is good.
At higher accretion rates, and therefore high burst rates, the residual
thermal energy from previous bursts becomes an important factor in
determining the ignition point, giving rise to the difference between the
\cite{cb00} and \cite{woos03} calculations (see the discussion and comparison in \S 5 of \citealt{woos03}). Fortunately, the range of
accretion rates spanned by \srcb\ is almost always $\lesssim 0.05\ \dot{m}_{\rm Edd}$.

\begin{deluxetable*}{lccccc}
\tablewidth{0pt}
\tablecaption{Properties of X-ray bursts from \srcb}
\tablehead{\colhead{} & \colhead{Burst start time}
& \colhead{Burst start time}
& \colhead{Peak flux} & \colhead{Fluence} &
\colhead{$\alpha$\tablenotemark{a}} \nl \colhead{} & \colhead{(UT)} &
\colhead{(MJD)} & \colhead{($10^{-9}\
\epcs$)} & \colhead{($10^{-6}\ \epc$)} & \colhead{} }
\startdata
 1 & 2002 Oct 15 09:55:37 & 52562.4136 & $163\pm3$ & $2.071\pm0.015$ & \nodata \\
 2 & 2002 Oct 17 07:19:24 & 52564.3051 & $177\pm3$ & $2.139\pm0.012$ & \nodata \\
 3 & 2002 Oct 18 04:25:20 & 52565.1843 & $179\pm3$ & $2.436\pm0.012$ & $148\pm3$ \\
 4 & 2002 Oct 19 10:14:33 & 52566.4268 & $177\pm3$ & $2.716\pm0.015$ & $167\pm3$ \\
 \enddata
 \tablenotetext{a}{The $\alpha$ values are calculated between pairs of
bursts, excluding the first two since we expect that intermediate bursts
may have been missed.}
\end{deluxetable*}

\section{Results}
\label{results}

 \subsection{The long-term mean accretion rate and corresponding distance lower limit}
 \label{distance}


We analysed all the available outbursts from
\srcb\ to 
calculate the time-averaged flux from
interval to interval.
The time-averaged accretion rate driven by angular momentum loss by gravitational radiation from the binary is \cite[]{bc01}
\begin{eqnarray}
  \dot{M}_{\rm GR} & \ga & 
      7\times10^{-12}\              M_\odot\,{\rm yr^{-1}}\ \left(\frac{M_C}{0.043M_\odot}\right)^2
  \nonumber \\ & & \times\ 
             \left(\frac{M_{\rm NS}}{1.4M_\odot}\right)^{2/3} \left(\frac{P_{\rm orb}}{2\ {\rm hr}}\right)^{-8/3} \label{mgr}
\end{eqnarray}
where $M_C$ is the minimum companion mass,
and $P_{\rm orb}$ the binary orbital period.
Assuming conservative mass transfer, we 
equated the measured time-averaged bolometric X-ray flux
$\left<F_X\right>$ and the predicted flux $(GM_{NS}/R_{NS})\dot{M}_{\rm
GR}/4\pi d^2$ to derive a lower limit on the distance of
\begin{eqnarray}
 d > 2.6\  {\rm kpc}\ \left(\frac{\left<F_X\right>}{10^{-10}\
\epcs}\right)^{-1/2}            \left(\frac{M_C}{0.043M_\odot}\right)\nonumber\\
     \times\      \left(\frac{P_{\rm orb}}{2\ {\rm hr}}\right)^{-4/3}
           \left({M_{\rm NS}\over 1.4\ M_\odot}\right)^{5/6}
           \left({R_{\rm NS}\over 10\ {\rm km}}\right)^{-1/2},
 \label{bddist}
\end{eqnarray}
where we choose the minimum companion mass $M_C=0.043M_\odot$, appropriate
for the mass function of \srcb\ of $3.78\times10^{-5}M_\odot$
\cite[]{chak98d} and assuming
a neutron star mass of $1.4M_\odot$.

\begin{figure}
 \epsscale{1.2}
 \plotone{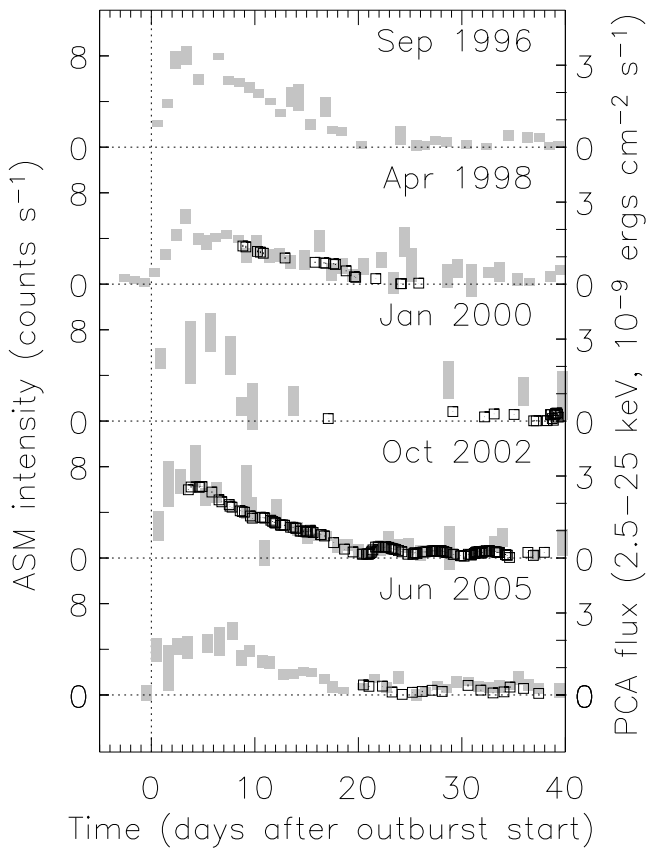}
 \figcaption[outburst-1808.eps]{\xte\/ ASM and PCA X-ray flux measurements
throughout the 
five outbursts of \srcb, from top to bottom: 1996 September, 1998 April,
2000 January, 2002 October and 2005 June. The gray boxes show the
$1\sigma$ confidence regions for the 1-d averaged ASM intensity estimates
(left-hand $y$-axis). The open squares show the PCA 2.5--25~keV flux
measurements (right-hand $y$-axis), with error bars indicating the
$1\sigma$
uncertainties (typically smaller than the symbols). The times for each
outburst have been shifted relative to the estimated outburst start times,
listed in Table \ref{outbursts}.
Note that additional proprietary PCA
observations (PI: Wijnands) were made in the first 20 days of the 2005
June outburst.
 \label{1808outbursts} }
\end{figure}

\srcb\ is the only accretion-powered MSP in which multiple outbursts have
been observed with \xte/PCA, although outbursts have been detected
retroactively in the long-term ASM flux histories of both \srcd\ and
\srcf\ \cite[]{markwardt02,gal05a}. We estimated the fluence for each
outburst of \srcb\ using public \xte\/ PCA and ASM measurements,  shown in Figure \ref{1808outbursts}.
Generally, the PCA observations (which provide the best measure of the
source flux) are triggered after the outburst has commenced, and often do
not cover the entire outburst, unlike the ASM measurements. For this reason we integrated the ASM intensities
over the parts of the outbursts not covered by the PCA measurements, and
determined a cross-calibration between the PCA flux and the ASM intensity,
as follows.
for each PCA flux measurement we selected the closest 1-day average of
the ASM 2--10~keV intensities, and performed a linear fit.
The fluences derived by this method, scaled to give the estimated
bolometric values, are listed in Table \ref{outbursts}.
We propagated the errors from the errors on individual ASM/PCA
measurements. We note that these values are roughly consistent with prior
estimates, e.g. for the 1998 April outburst \cite{gilfanov98} estimated
$4.2\times10^{-3}\ \epc$ (3--150~keV), whereas we estimate
$(5.2\pm0.5)\times10^{-3}\ \epc$ (extrapolated 0.1--200~keV).

The derived fluences were constant to within the errors, with variation
only at the $2.6\sigma$ level. However, the interval between outbursts
increased after each of the the first three, from 1.58~yr between the 1996
September and 1998 April outbursts, to 2.8~yr between the 2000~January and
2002~October outbursts.
Although the interval to the 2005 June outburst was slightly less than
the previous interval, the integrated fluence 
was the lowest of all the outbursts.
As a result, the time-averaged X-ray flux dropped
by 40\% between 1996 and 2002. Since the derived distance limit depends on
the mean flux only to the $-1/2$ power (equation \ref{bddist}), the
derived distance limits varied only by 30\%, up to a maximum of 
3.4~kpc
(Table \ref{outbursts}).
The mean X-ray flux, based on the four outbursts observed since 1996
September, was $8\times10^{-11}\ \epcs$, which (for a distance of
3.6~kpc) corresponds to approximately $10^{-11}\ M_\odot\,{\rm yr}^{-1}$.

\begin{figure}
 \epsscale{1.2}
 \plotone{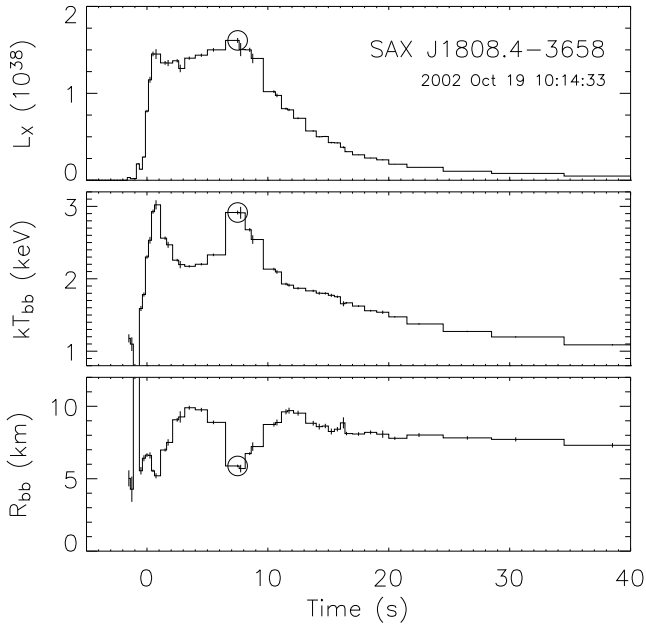}
 \figcaption[]{The fourth, and most intense, burst observed from \srcb\ by
\xte\/ during the 2002 October
outburst. {\it Top panel}\/ The estimated bolometric flux is plotted as a
histogram, with error bars indicating the $1\sigma$ uncertainties; at the
burst peak, only short intervals of data were obtained
due to telemetry buffer overruns. The distance is assumed to be 3.6~kpc.
{\it Middle panel}\/ 
The 
blackbody temperature $kT_{\rm bb}$.
{\it Bottom panel}\/
The blackbody radius $R_{\rm bb}$. Note the anticorrelated behavior of 
$kT_{\rm bb}$ and $R_{\rm bb}$ during the burst peak.
The circle marks the time when the peak burst flux was achieved.
\label{1808bursts} }
\end{figure}

 \subsection{Comparison of burst properties to ignition models}
 \label{bursters}

We now compare the properties of thermonuclear bursts from \srcb\ with predictions from theoretical ignition models. 
Observed properties of the bursts are listed in Table 2. The bursts were quite homogeneous, with standard devation of the fluences of only about 12\%.
The bursts were all observed within a 100~hr interval just after the
outburst peak, and the last three were separated by 21.1 and 29.8~hr,
respectively. This increase in burst interval was observed as the
2.5--25~keV flux
decreased from 2.3 to $1.9\times10^{-9}\ \epcs$.
Gaps between the observations with bursts make it possible that
intermediate events were missed. Assuming that was not the case, the
estimated $\alpha$-values for the second two intervals were 
$\alpha=148\pm3$, $167\pm3$
respectively.  These values, along with the fast rise times
($\approx0.5$~s) 
suggest pure He fuel. This is supported by the lightcurves of the X-ray bursts, an example of which is shown in Figure \ref{1808bursts}. The long $\approx 10$ second period of constant (presumably Eddington) luminosity followed by a decay over $\approx 10$ seconds is remarkably similar to the results of \cite{woos03} for bursts which ignite helium in a pure helium environment (compare with their Fig.~26).
As we will show, the recurrence times and energetics of the  bursts also suggest that helium ignition is occurring in a pure helium layer beneath steady hydrogen burning shell.

\subsubsection{Pure-helium triggered X-ray bursts}

Before presenting our detailed fits to the ignition models, we first make some analytic arguments which help to understand the observed properties of the bursts and the results of our fits. First, consider the burst energetics. The observed values of $\alpha\approx 150$ imply that the nuclear energy release per gram is $Q_{\rm nuc}=c^2z/\alpha\approx 2$ MeV per nucleon, where we take the gravitational redshift $1+z=1.31$. Using the relation $Q_{\rm nuc}=1.6+4\langle X\rangle$ MeV per nucleon gives an average hydrogen mass fraction $\langle X\rangle\approx 0.1$ in the burning layer immediately prior to ignition.

The low mean hydrogen fraction at ignition comes about because hot CNO
hydrogen burning leads to steady depletion of the hydrogen as the H/He
layer accumulates. The time to burn all the hydrogen in a given fluid
element depends on the initial hydrogen fraction $X_0$ and the mass
fraction of CNO elements $Z_{\rm CNO}$. In the observer's frame, this time
is $t_{\rm burn}=38\ {\rm hrs}\ X_0\ ({Z_{\rm CNO}/0.01})^{-1}(1+z)/1.31$.
The bursts from \srcb\ have recurrence times $\Delta t\approx 1$ day and
$\langle X\rangle\approx 0.1$. This immediately constrains $Z_{\rm CNO}$
and $X_0$: only particular combinations of these two parameters will lead
to an average hydrogen mass fraction $\approx 0.1$ for the observed recurrence time of $\approx 1$ day. The average hydrogen fraction in the layer at ignition is $\langle X\rangle=(X_0/2)(t_{\rm burn}/\Delta t)$, where we have used the fact that the hydrogen fraction decreases linearly with depth \citep{cb00}. Setting $\langle X\rangle$ and $\Delta t$ to the observed values, we find 
\begin{eqnarray}
 X_0 & = &  
0.36\ \left({Z_{\rm CNO}\over 0.01}\right)^{1/2}\left({\Delta t\over 1\ {\rm day}}\right)^{1/2}
 \nonumber \\ & & \times\ 
\left({\langle X\rangle\over 0.1}\right)^{1/2}\left({1+z\over 1.31}\right)^{-1/2}
\label{eq:zxcorr}
\end{eqnarray}
is the required relation between $X_0$ and $Z_{\rm CNO}$ to reproduce the
measured values of $\Delta t$ and $\langle X\rangle$\footnote{Although not
shown explicitly, the normalization of equation \ref{eq:zxcorr} is quite
sensitive to the assumed redshift through the factor $<X>$, since for a
given $\alpha$, $Q_{\rm nuc}\propto z$. For example, a larger choice for
the redshift of $z=0.42$, which gives $\langle X\rangle=0.25$, shifts the
relation in equation \ref{eq:zxcorr} so that it admits both solar values
for $X_0$ and $Z_{\rm CNO}$. However, as we describe in \S3.2.2, our
detailed fits, which account for the variations in burst properties from
burst to burst, constrain the relation between $X_0$ and $Z_{\rm CNO}$ to
be approximately that given by  equation \ref{eq:zxcorr}, or in other
words imply that $z\approx 0.3$.  The resulting constraints on mass and
radius are described in \S3.2.2 (see discussion below eq.~13).}.

To go further, we need to understand the ignition column depth. An example
of a temperature profile in the layer for $\Delta t>t_{\rm burn}$ is given
in Figure 1(b) of Cumming \& Bildsten (2000). The temperature rises
through the hydrogen burning layer as the hydrogen burns, but the layer of
pure helium which accumulates beneath it is close to isothermal. Therefore
the temperature at the base of the hydrogen burning layer is a good
approximation of the ignition temperature. The temperature profile in the
hydrogen burning layer is given by $dT/dy=3F\kappa/4acT^3$, where $\kappa$
is the opacity, and $y$  is the column depth (units of ${\rm g\
cm^{-2}}$). The flux is $F=\epsilon_H(y_d-y)$ where $y_d=\dot m X_0
E_H/\epsilon_H$ is the depletion depth for hydrogen. The energy release in
the hot CNO cycle is $E_H=6.0\times 10^{18}\ {\rm erg\ g^{-1}}$; the
heating rate from the hot CNO cycle is $\epsilon_H=5.8\times 10^{13} \
{\rm erg\ g^{-1}\ s^{-1}}\ (Z_{\rm CNO}/0.01)$. Integrating from the
surface, we find the temperature at the base of the hydrogen layer is
given by\footnote{The base temperature $T_b$ has some
dependence on composition via the opacity. However, this is a small
effect. Varying the hydrogen fraction or metallicity from zero to solar
values, we find that the resulting changes in opacity are $\lesssim 50$\%,
and the changes in temperature 
(eq.~\ref{eq:tb}) and inferred accretion rate (eq.~\ref{eq:acc}) smaller
still because of the weak dependences $T_b\propto\kappa^{1/4}$ and $\dot
m\propto \kappa^{-9/22}$.}
\begin{equation}
T_b^4={3\kappa\over 2ac}{\dot m^2X_0^2E_H^2\over \epsilon_H}\propto {X_0^2\over Z_{\rm CNO}}.
\label{eq:tb}
\end{equation}
But we have already seen that the energetics fix the ratio $X_0^2/Z_{\rm CNO}$ (eq.~[\ref{eq:zxcorr}]), and so the temperature $T_b$ is independent of $X_0$ or $Z_{\rm CNO}$,
\begin{eqnarray}\label{eq:Tb}
T_b&=&\left({3\kappa E_H\dot m^2\langle X\rangle\Delta t\over ac (1+z)}\right)^{1/4}\nonumber\\
&=&2.1\times 10^8\ {\rm K}\ \left({\dot m\over 0.1\dot m_{\rm Edd}}\right)^{1/2}\left({\langle X\rangle\over 0.1}\right)^{1/4}\left({\Delta t\over 24\ {\rm h}}\right)^{1/4},
\end{eqnarray}
where we set $\kappa=0.05\ {\rm cm^2\ g^{-1}}$ as a typical value for opacity and $z=0.31$.

Next we assume that the helium layer is isothermal, so that $T_b$ is also the temperature at the helium ignition depth. It can be shown that this is a good assumption as long as the flux heating the helium layer from below is $Q_b\lesssim E_H\langle X\rangle\approx 0.6\ {\rm MeV\ per\ nucleon}$ for $\langle X\rangle$ of $0.1$, which is satisfied for the expected values of $Q_b\sim 0.1$ MeV per nucleon (e.g.~\citealt{brown00}). A good fit to the helium ignition curve of \cite{cb00} is $y_b=10^{10}\ {\rm g\ cm^{-2}}\ (T_b/1.1\times 10^8\ {\rm K})^{-9}$ (for $y_b>2\times 10^8\ {\rm g\ cm^{-2}}$). Substituting equation (\ref{eq:Tb}) for $T_b$ into this fit, and solving for $\dot m=y_b(1+z)/\Delta t$ gives
\begin{equation}\label{eq:acc}
\dot m=0.057\ \dot m_{\rm Edd}\ \left({\Delta t\over 24\ {\rm h}}\right)^{-13/22}\left({\langle X\rangle\over 0.1}\right)^{-9/22},
\end{equation}
where we again set $\kappa=0.05\ {\rm cm^2\ g^{-1}}$.

To summarize, we have found that given a sequence of pure helium ignition bursts with measured $\Delta t$ and $\langle X\rangle$ (or equivalently $\Delta t$ and $\alpha$), $X_0$ and $Z_{\rm CNO}$ must be related according to equation (\ref{eq:zxcorr}), and the accretion rate is determined by equation (\ref{eq:acc}) independent of $X_0$ or $Z_{\rm CNO}$. Writing $F_X=\dot mQ_{\rm grav}(R/d)^2/(1+z)$ where $Q_{\rm grav}=c^2z/(1+z)\approx GM/R$ is the gravitational energy release per gram, and solving for the distance, we find
\begin{eqnarray}
d & = & 4.15\ R_6\ \left({F_X\over 5\times 10^{-9}\ {\rm erg\ cm^{-2}\ s^{-1}}}\right)^{-1}
\nonumber \\ & & \times\ 
\left({\Delta t\over 24\ {\rm h}}\right)^{-13/44}\left({\langle X\rangle\over 0.1}\right)^{-9/44}\ {\rm kpc}
\end{eqnarray}
which is independent of $X_0$ and $Z_{\rm CNO}$.

\begin{figure}
 \epsscale{1.2}
 \plotone{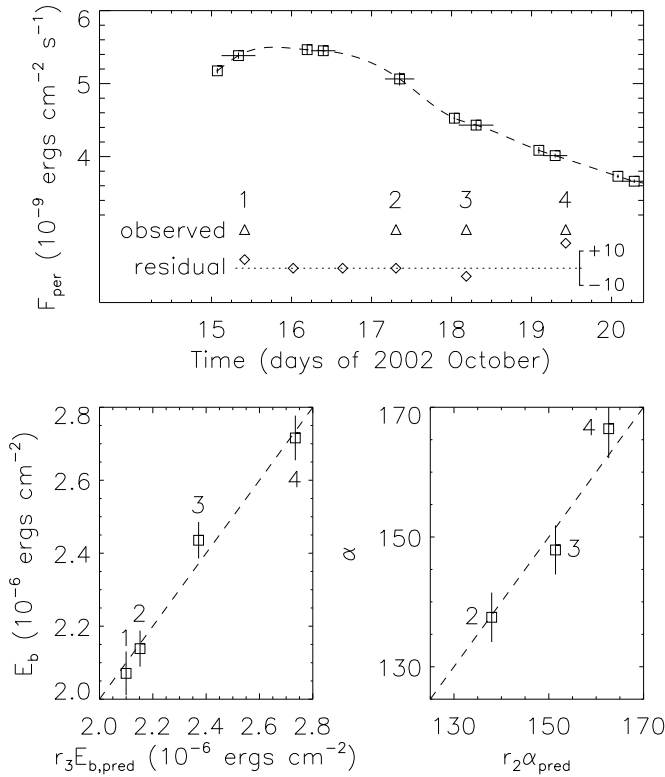}
 \figcaption[]{Comparison of model results and observed burst properties
for \srcb, for 
$Q_b=0.325$~MeV/nucleon,
$Z_{\rm CNO}=0.02$, $X_0=0.54$ ($\chi^2=6.2$
for 6 dof). The top panel shows the inferred bolometric flux ({\it open
squares}) and the interpolation used to estimate the integrated $\dot{m}$
between the bursts ({\it dashed line}). The open triangles are the
observed burst times, and the open diamonds 
indicate the error (observed-predicted) in the model burst times
(relative to $t_2$). {\it Lower left} The observed and predicted fluence,
the latter rescaled by $r_2$; the dashed line indicates a 1:1
correspondence. {\it Lower right} The observed and predicted
$\alpha$-values, the latter rescaled by $r_3$. The measurement for burst 2
is a mean assuming that two bursts were missed between burst 1 and 2.
 \label{bestresult} }
\end{figure}

\subsubsection{Fitting procedure and results}

We can improve on the estimates of the previous section by firstly fitting
the observations to the full ignition models (which, for example, include
accurate opacity values as a function of depth in the layer), and also by
fitting for the change of burst recurrence time with changing accretion
rate during the outburst. We generated sequences of bursts based on the inferred $\dot{m}$ history
of \srcb\ during the 2002 outburst as follows. First, for a given set of
model parameters
($Q_b$, $Z_{\rm CNO}$ and $X_0$) we identified the
values of $\dot{m}$ for which the recurrence time was closest to that of
the third and fourth bursts (assuming no intermediate bursts occurred),
$\Delta t_{2}=21.10$~hr and $\Delta t_{3}=29.82$~hr. From these 
$\dot{m}_2$, $\dot{m}_3$ values we calculated the ratio of the persistent
fluxes to the inferred $\dot{m}$, i.e.
\begin{equation}
  r_1 \equiv \left< \frac{\dot{m}_i}{c_{\rm bol}F_{p,i}} \right>
  \label{r1first}
\end{equation}
where
$F_{p,i}$ is the mean 2.5--25~keV persistent flux
(in units of $10^{-9}\ \epcs$, from 
cubic spline
interpolation) between burst $(i-1)$ and $i$. The value of $r_1$ will
obviously depend upon the distance to the source, as well as the neutron
star mass and radius, but at this stage we assume no particular values for
these parameters, only that the relationship between $F_p$ and $\dot{m}$
is constant over the interval where the bursts occurred.

We converted the interpolated flux history of \srcb\ to an $\dot{m}$
history by scaling by $r_1$, and generated sequences of bursts forward
in time from the first burst using the ignition model. 
We then matched simulated bursts predicted to occur at approximately the
same time as the observed bursts,
and compared their properties.
Specifically, we compared the model predicted variation of the fluence and
$\alpha$-values by calculating two additional ratios
$ r_2  \equiv  \left< \frac{\alpha}{\alpha_{\rm pred}}\right>$
(where $\alpha_{\rm pred}=290/Q_{\rm nuc,pred}$ is the predicted value from
the model, 
and
$\alpha = F_p c_{\rm bol} \Delta t/E_b $
is the observed value),
and
$ r_3 \equiv \left< \frac{E_b/10^{-9}}{L_{b,{\rm pred}}} \right>$
(where $L_{b,{\rm pred}}$ is the predicted burst luminosity in $10^{39}\
\eps$).
We scaled $\alpha_{\rm pred}$ and $L_{b,{\rm pred}}$ by the factors
$r_2$ and $r_3$, respectively, in order to compare with the observed
values. 
Effectively, we try to match the observed variation of $E_b$ and $\alpha$
with $\dot{m}$, under the assumption that the average offset between the
observed and predicted values is a function of the system parameters (see
\S3.2.3).
In order to quantify the agreement with the observed values, we
calculated an overall $\chi^2$ for the simulated burst properties, 
including predicted burst times for the second through fourth bursts;
\begin{eqnarray}\label{eq:chi2}
\chi^2 & = & \sum\left(\frac{\alpha-r_2\alpha_{\rm pred}}{\sigma_\alpha}\right)^2
       +\sum\left(\frac{E_b-r_3L_{b,{\rm pred}}}{\sigma_{E_b}}\right)^2
\nonumber \\ & & 
       +\sum\left(\frac{t_i-t_{i,{\rm pred}}}{\sigma_t}\right)^2
\end{eqnarray}
where we adopted a representative uncertainty for the burst arrival times
of $\sigma_t=10$~min chosen so that the contribution of the last term in equation (\ref{eq:chi2}) was comparable to the other terms in the sum (the burst start times are actually measured to a
precision of less than a second).

We carried out the simulation for a range of input parameters defined by
grids of
$Q_b$ between 0.175--0.5~MeV/nucleon (in steps of 0.025),
$Z_{\rm CNO}=0.006$--0.034 (steps of 0.002) and
$X_0=0.28$--0.74 (steps of 0.02), and calculated the $\chi^2$
for each grid point.
The burst times were reproduced well over a range of the input parameters,
with an error as small as just 0.2~hr rms.  The variation in $E_b$ and
$\alpha$ was less well fitted, 
giving an overall $\chi_{\rm min}^2=49$.  With three burst arrival
times, three $\alpha$ values and four measured fluences, and three
parameters in the fit
($Q_b$, $Z_{\rm CNO}$ and $X_0$) the number of
degrees of freedom is 6, so that the best fit is not statistically
acceptable.

\catlabel\ estimated the errors on the fluence for the bursts by
propagating the uncertainties in the measured flux in each time bin.
However, during the 6--8~s PRE episode in each burst the X-ray flux was
between $1.6\times10^{-7}$ and
$1.8\times10^{-7}\ \epcs$, so that the onboard telemetry buffer was filled
more rapidly than it could be read out (once per second; Fig.
\ref{1808bursts}).  As a result, we have incomplete information about the
flux at the peak of the bursts, which makes it likely that the uncertainty
on the fluences ($\approx0.6$\% on average) was underestimated.
The variation of the flux at the peak of the burst of $\sim10^{-8}\
\epcs$, combined with the typical observational duty cycle there (due to
the gaps) of $\approx30$\% suggests that the actual fluence uncertainty is
around a factor of 2 higher.  By increasing the error on the fluences by
this factor, and propagating the uncertainties for the $\alpha$
calculations, we obtained a $\chi^2$ value which was still significantly
larger than the number of degrees of freedom. Increasing the errors on the
fluence by an additional factor of two was sufficient to achieve a minimum
$\chi^2$ of 6.2, comparable to the number of degrees of freedom. 
We compare the measured burst properties and the best set of model results
graphically in Fig. \ref{bestresult}.

\begin{figure}
 \epsscale{1.2}
 \plotone{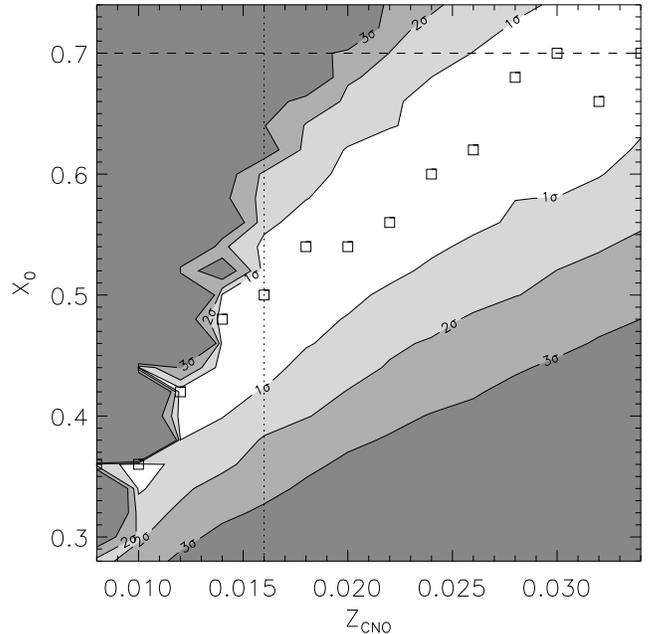}
 \figcaption[]{Contours of $\chi^2$ as a function of $Z_{\rm CNO}$, $X_0$
for the ignition model fits to the 2002 October burst sequence observed
from \srcb, described in \S\ref{bursters}.  The dotted line corresponds to
solar metallicity, while the dashed line indicates solar H-fraction. The
best solution for each grid value of $Z_{\rm CNO}$ is indicated by the open
square.
 \label{chigrid} }
\end{figure}

We also estimated from the grid results the confidence region for each of
the three parameters. 
The lowest values of $\chi^2$ were consistently achieved for 
$Q_b=0.3$~MeV/nucleon.
In order to illustrate graphically the combined confidence region
for the other two parameters, we flattened the grid along the 
$Q_b$ axis to plot Figure \ref{chigrid}.
We adopt as a confidence level the probability
corresponding to a $1\sigma$ deviation, for which the customary confidence
region has 
$\Delta \chi^2=\chi^2-\chi^2_{\rm min}<2.30$ \cite[for two
parameters of interest; e.g.][]{lmb76}.
We found the best agreement of the measured burst properties to the
ignition model within a diagonal strip in $Z_{\rm CNO}$-$X_0$ space. This correlation is in good agreement with equation (\ref{eq:zxcorr}) which follows from consideration of the burst energetics. We could not obtain an acceptable model fit for
$X_0\la0.3$. The best solution for solar $Z_{\rm CNO}=0.016$ required 
$X_0=0.5$; in contrast, for a solar $X_0=0.7$, the corresponding $Z_{\rm CNO}$
was substantially above the solar value at 0.03.

As a consistency check we compared the predicted times of bursts which
were not detected in the \xte\/ observations, with the time intervals in
which \xte\/ was observing the source. We predicted two additional bursts
on October 16 (Fig. \ref{bestresult}), between the first two bursts
detected by \xte; none of the model runs which agreed well with the
observations (i.e. $\Delta\chi^2\leq3.51$) predicted times for these
bursts which fell within the observations on that day.
We also examined the seventh burst in the series, predicted to follow the
last burst detected by \xte\/ on October 19.  The set of models that
agreed well with the observations predicted times for the seventh burst
between MJD~52568.398 and 52568.601 (October 21); just five predicted
times fell within the time interval for one of the observations on that
day, compared to 273 which fell within the gap. Thus, we estimate the
probability of having observed this burst as 0.018, which is consistent
with the lack of detection by \xte.

The ignition models also allow us to estimate the mean H-fraction at
ignition for each of the bursts. Over the $1\sigma$ confidence region in
$Z_{\rm CNO}$-$X_0$ space, the model-predicted mean hydrogen fraction at
ignition $\left<X\right>$ for each burst varied by $\sim10$\%. For the
four bursts observed by \xte, we find $\left<X\right>=0.191\pm0.016$,
$0.176\pm0.015$, $0.133\pm0.011$ and $0.095\pm0.008$, respectively. The
decrease in $\left<X\right>$ is consistent with the steadily increasing
burst recurrence time, which allows more of the accreted H to be removed
by steady burning.

\begin{figure}
 \epsscale{1.2}
 \plotone{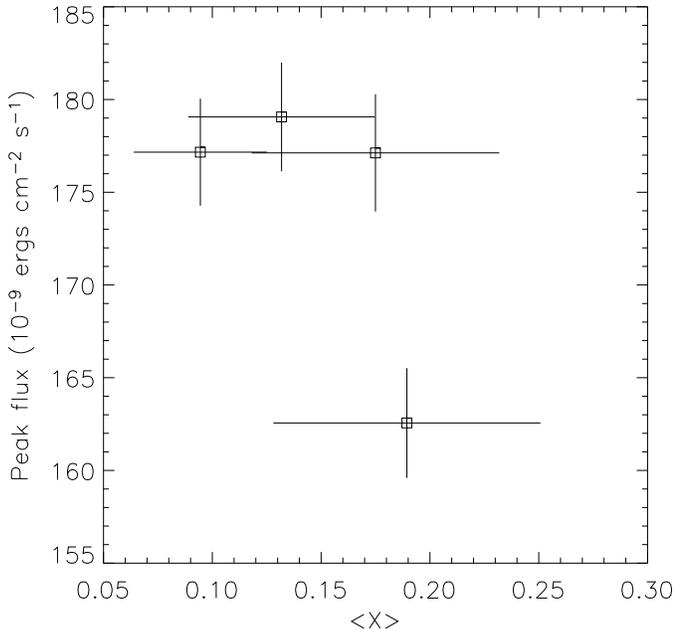}
 \figcaption[]{Peak burst flux as a function of the mean hydrogen mass
fraction $\left<X\right>$ inferred from the ignition model, for the four
bursts observed by \xte\/ from \srcb\ during the 2002 October outburst.
Error bars on the flux indicate the $1\sigma$ uncertainties; for
$\left<X\right>$, the range spanned by the set of model results for which
the $\chi^2$ values are within the $1\sigma$ confidence limits in Fig.
\ref{chigrid}.
 \label{pflux} }
\end{figure}

\subsubsection{Distance estimate}

Based on our simulation results, we attempted to constrain the distance
to \srcb.  The accretion rate (per unit area) is
\begin{eqnarray}
  \dot{m} & = & \left(\frac{d}{R}\right)^2\,
      \frac{F_{\rm p} c_{\rm bol}(1+z)}
      {Q_{\rm grav}} \nonumber \\
   & = & 6.7\times10^3\ \left(\frac{F_p c_{\rm bol}}{10^{-9}\ \epcs}\right)
            \left(\frac{d}{10\ {\rm kpc}}\right)^2
\nonumber \\ & & \times\  
            \left(\frac{M_{\rm NS}}{1.4\ M_\odot}\right)^{-1} 
                      \left(\frac{1+z}{1.31}\right)
            \left(\frac{R_{\rm NS}}{10\ {\rm km}}\right)^{-1}\
            {\rm g\,cm^{-2}\,s^{-1}}
\label{mdot}
\end{eqnarray}
where  
$d$ is the distance to \srcb, and
$Q_{\rm grav}=c^2z/(1+z)\approx GM_{\rm NS}/R_{\rm NS}$ is the energy
released per nucleon during accretion.
Then
\begin{eqnarray}
  r_1 & = & 0.0762
			\left(\frac{d}{10\ {\rm kpc}}\right)^2
			\left(\frac{M_{\rm NS}}{1.4M_\odot}\right)^{-1}
\nonumber \\ & & \times\ 
			\left(\frac{1+z}{1.31}\right)
			\left(\frac{R_{\rm NS}}{10\ {\rm km}}\right)^{-1}
  \label{r1}
\end{eqnarray}
Most of the terms in equation \ref{r1} are expected to be of order unity.
Thus,
we adopt $d_1=10\sqrt{r_1/0.0762}$~kpc as a rough estimate of the distance;
we found a
range of 
3.1--3.5~kpc over the set of model parameters $Q_b$, $Z_{\rm CNO}$ and
$X_0$ for which the
$\chi^2$ value was within the $1\sigma$ limit for 3 parameters
\cite[$\Delta\chi^2=3.51$;][]{lmb76}.

Similarly, we calculate
\begin{eqnarray}
				\nonumber \\
r_3 & = & 82.6
			\left(\frac{d}{10\ {\rm kpc}}\right)^{-2}
			\left(\frac{R_{\rm NS}}{10\ {\rm km}}\right)^{2}
\nonumber \\ & & \times\
			\left(\frac{1+z}{1.31}\right)^{-1}
			\left(\frac{Q_{\rm nuc}}{Q_{\rm nuc,pred}}\right)
  \label{r3}
\end{eqnarray}
so that we obtain a second estimate of the distance
$d_3=10\sqrt{82.6/r_3}$~kpc based on the observed and predicted fluences.
Over the $1\sigma$ confidence range of ignition model parameter space we
find a distance range of 3.6--3.8~kpc.
That the ranges of $d_1$ and $d_3$ do not coincide indicates that at least
one of the terms in equations \ref{r1} and \ref{r3} deviates significantly
from 1. Because the $R_{\rm NS}$ term is the only one common to both
expressions, and it appears in both distance expressions with exponents of
the same sign, we infer that no single term can give rise to the
disagreement.

The calculated values of $r_2$ can give an indication of which parameters
may vary from our estimated values. We derive
\begin{eqnarray}
  r_2 
      & = & 0.668 
			\left(\frac{M}{1.4M_\odot}\right)
			\left(\frac{R}{10\ {\rm km}}\right)^{-1}
\nonumber \\ & & \times\
		\left(\frac{Q_{\rm nuc}}{Q_{\rm nuc,pred}}\right)^{-1}
  \label{r2}
\end{eqnarray}
and we found over the $1-\sigma$ confidence region $r_2/0.668$ varies
between 1.56 and 1.79. This suggests that the neutron-star mass must
exceed $1.4M_\odot$, or that the radius must be $<10$~km or that $Q_{\rm
nuc}<Q_{\rm nuc,pred}$ (or some combination of the three). A larger
$M_{\rm NS}>1.4M_\odot$ would explain the lower value of $r_1$ (equation
\ref{r1}) leading to the distance estimate $d_1<d_3$; similarly, $Q_{\rm
nuc}<Q_{\rm nuc,pred}$ would lead to an inflated value of $d_3$. On the
other hand, a smaller $R_{\rm NS}<10$~km would drive both $d_1$ and $d_3$
to even greater disagreement. Thus, we conclude that the likely
explanation for the disagreement between $d_1$ and $d_3$ is that $M_{\rm
NS}>1.4M_\odot$ and $Q_{\rm nuc}<Q_{\rm nuc,pred}$, and that the distance
range is between 3.1--3.8~kpc.

 \subsection{Distances from thermonuclear burst peak fluxes}
 \label{pre}

Presently, the most 
commonly-used method to determine the distance to a bursting LXMB
is to measure the peak flux
of thermonuclear bursts which exhibit photospheric radius-expansion.
Such bursts exhibit a localized maximum in the fitted
blackbody radius in the first few seconds of the burst, accompanied by a
local minimum in the blackbody temperature.
Several authors have shown that the peak flux of radius-expansion
bursts for sources with known distances is a rough standard candle; e.g.
\cite{kuul03a} found that a sample of radius-expansion bursts from globular
cluster LMXBs almost all reached a common peak luminosity of
$(3.79\pm0.15)\times10^{38}\ \eps$. This empirical value is somewhat
higher than predicted by theory \cite[]{lew93}
\begin{eqnarray}  
L_{\rm Edd,\infty} & = & \frac{8\pi G m_{\rm p} M_{\rm
NS} c
  [1+(\alpha_{\rm T}T_{\rm e})^{0.86}]} {\sigma_{\rm T}(1+X)[1+z(R)]}
       \nonumber \\
  & = & 2.7\times10^{38} \left(\frac{M_{\rm NS}}{1.4M_\odot}\right)
 \frac{1+(\alpha_{\rm T}T_{\rm e})^{0.86}}{(1+X)}
\nonumber \\ & & \times\
    \left[\frac{1+z(R)}{1.31}\right]^{-1} 
\eps
  \label{ledd}
\end{eqnarray}
where
$T_{\rm e}$ is the
effective temperature of the atmosphere, $\alpha_{\rm T}$ is a coefficient
parametrizing the temperature dependence of the electron scattering
opacity \cite[$\simeq 2.2\times10^{-9}$~K$^{-1}$;][]{lew93} and $X$ is the
mass fraction of hydrogen in the atmosphere at ignition.
The final factor in
square brackets represents the gravitational redshift 
at the photosphere $1+z(R)=(1-2GM_{\rm NS}/R
c^2)^{-1/2}$. The photosphere may be elevated significantly above the NS
surface (i.e. $R\ge R_{\rm NS}$), in which case this is a small correction.

All four bursts from \srcb\ observed by \xte\/ during the October 2002
outburst showed unambiguous indications of radius expansion
\cite[see also][]{chak03a}.
These bursts all 
exhibited a local maximum in fitted blackbody radius and local minimum in
blackbody temperature (Fig. \ref{1808bursts}) immediately prior to the
time at which maximum flux was reached. The average peak
flux was $(174\pm8)\times10^{-9}\ \epcs$, indicating a distance of
$d=2.8$ (3.6)~kpc for $X=0.7$ (0.0) in equation \ref{ledd}
(see also \catlabel).
These values are approximately the same as those derived by \cite{zand01}
from the bursts observed first by \sax\ 
during the 1996 September outburst.
The corresponding distance assuming the empirical value of \cite{kuul03a}
is 4.3~kpc. The best agreement with our distance estimate from \S 3.1 is
for $X\approx 0$ at the photosphere. Without a detailed model of the
expansion of the outer layers, it is not possible to say how the
composition at the photosphere relates to the mean hydrogen fraction in
the layer. Nonetheless, we note that there is a rough anticorrelation
between the peak flux of the burst and $\langle X\rangle$ from our
ignition model fits. This is shown in Figure \ref{pflux}.

\section{Discussion}
\label{disc}

We have 
analysed the 
X-ray flux evolution during five outbursts of \srcb, as
well as
the sequence of thermonuclear (type I) X-ray bursts observed during the
2002 October outburst. The large $\alpha$-values estimated for the bursts
indicates low
mean H-fraction $\left<X\right>\approx0.1$ at ignition, 
which analytically
implies a common constraint on the accreted H-fraction $X_0$
and metallicity $Z_{\rm CNO}$.
We confirmed this result by comparing the measured burst properties and
recurrence times
with predictions by the burst ignition model of \cite{cb00}. We found
that the inferred burst recurrence times 
were sufficiently long that the
helium ignition occurs in
a pure helium environment, the accreted hydrogen having completely burned away
between bursts by the hot CNO cycle.
While the model comparison admits a range in both $Z_{\rm CNO}$ and $X_0$,
it excludes solar values for both.  If the metallicity is approximately
solar, then the hydrogen fraction in the accreted material $X_0$ must be
significantly sub-solar; on the other hand, if $X_0$ is approximately at
the solar value, then $Z_{\rm CNO}$ must be significantly above the solar
value.
Many stars in the Galaxy have metallicities twice solar, and the progenitor of \srcb\ may have been a 2--3~$M_\odot$
star. The metallicities of other stars in the neighborhood may provide
circumstantial evidence in support of super-solar values for \srcb,
although this would be far from conclusive.

We used our fits to the burst models to constrain the distance to the
source. Our analytic estimates suggest that the local accretion rate onto
the star, and therefore the distance, is determined independently of the
accreted composition in this burst regime. Our model fits support this,
giving a distance range of $3.1$--$3.8\ {\rm kpc}$ across the grid of the parameters $X_0$, $Z_{\rm CNO}$, and $Q_b$. We also used fact that the bursts showed photospheric radius expansion to constrain the distance, assuming that the peak burst flux was equal to the local Eddington luminosity. This gives a distance of $3.6\ {\rm kpc}$ for a pure helium photosphere.

Finally, we obtained a constraint on the distance based on measurements of
the fluence and recurrence time for each of the
outbursts observed since 1996 September.
The fluences were approximately constant between outbursts, but the
increase in the interval between outbursts indicates that the long-term
accretion rate is steadily increasing.
We derived a lower limit on the distance based on the
assumption that the long-term mass transfer rate is driven by
gravitational radiation in the binary.
Since the long-term average flux is decreasing, this limit has increased
with each successive outburst, to a maximum of 3.4~kpc for the interval
prior to the 2005 June outburst.
This
limit excludes all but the largest values in the range derived
from the peak burst flux. By combining this limit and the estimated range
based on the peak burst flux, we derive a probable distance range of
3.4--3.6~kpc.

It is reassuring that the distance limits derived from the long-term
flux history are consistent with our other measurements, despite the 
variations in $\dot{m}$.
This is
particularly relevant to the transient accretion-powered pulsars in which
no bursts are observed. That the long-term averaged $\dot{m}$ varies at
all in \srcb\ is unexpected, since the timescale for the variation is much
shorter than the characteristic evolution time of these systems (of order
$10^9$~yr). 
The similarity of the lower distance limit derived from the long-term flux
and the estimated upper limits from the bursts suggests that the mass-transfer
rate has already reached $\approx\dot{M}_{\rm GR}$, and thus could be
expected to remain at least at that level for successive outbursts.
Alternatively, the observed variation may be part of a longer-term cycle,
and $\dot{m}$ may recover in the future. 

As a caveat regarding the burst fitting procedure, we point out that
we neglect
possible systematic contributions,
which are 
difficult to quantify. The most
dramatic potential effect would be if we have missed every second burst in
the regular Earth occultations, so
that the actual burst rate is twice what we infer. However, we can rule
this out fairly confidently, since the required $\dot{m}$ would be a
factor of $\sim2$ higher, in which case (unless 
the persistent flux is preferentially beamed away from the line of sight) the distance
to the source would be around 40\% greater. This would be significantly
outside the probable range derived here or any range quoted in the
literature \cite[e.g.][]{zand01}.
\cite{wang01} derived distance contours based on the measured $A_V$ of the
optical counterpart, and estimated an inclination of $\cos i=0.8$ for a
distance of 3~kpc.  For our estimated lower distance limit of 3.4~kpc, we
infer $\cos i>0.8$, so that $i\la30^\circ$.

Our best-fitting models give a value $Q_b\approx 0.3$ MeV per nucleon for
the flux from deep in the star that heats the accumulating fuel layer. For
an accretion rate $\approx 0.06\ \dot m_{\rm Edd}$ (eq.~\ref{eq:acc}), the
corresponding luminosity is $4\pi R^2\dot mQ_b\approx 2\times 10^{34}\
{\rm ergs\ s^{-1}}$. This value is three orders of magnitude larger than
the quiescent luminosity of SAX~J1808.4$-$3658. \cite{campana02}
modeled the quiescent spectrum as a blackbody plus power law component,
and obtained an upper limit to the blackbody flux of $2\times 10^{-15}\
{\rm ergs\ cm^{-2}\ s^{-1}}$. For a distance of $3.5\ {\rm kpc}$, the
implied X-ray luminosity is $<3\times 10^{30}\ {\rm ergs\ s^{-1}}$. For a
spectral model which included a hydrogen atmosphere model rather than
blackbody, the upper limit was a factor of two lower. This large
discrepancy between the outwards flux deduced from X-ray burst properties
and the flux measured in quiescence is similar to the case of KS~1731$-$260.
In that case, the observed quiescent luminosity is $\approx 3\times
10^{33}\ {\rm ergs\ s^{-1}}$ (for an assumed distance of 8 kpc; Rutledge et
al.~2002). However, KS~1731$-$260 showed a superburst during outburst, for
which a luminosity 20 times greater, $\gtrsim 6\times 10^{34}\ {\rm ergs\
s^{-1}}$, is required in order to achieve unstable carbon ignition at the
correct depth (Figure 20 of Cumming et al.~2006). The meaning of these
discrepancies is not yet clear, but may indicate some additional heating
of the deep ocean that is not included in current models of accreting
neutron stars.

Our new distance estimate for SAX~J1808.4$-$3658 eases the constraints on
neutron star interior models from the quiescent luminosity.  \cite{campana02}
noted that the low quiescent luminosity implies that the
neutron star core is cold, presumably indicating a massive neutron star
core with fast neutrino emission. Yakovlev et al. (2004) further found
that only models with hyperons in the core could explain the low
luminosity. However, they assumed a distance of 2.5~kpc, whereas our new
distance of $\approx 3.5$ kpc implies a luminosity two times greater than
that assumed by Yakovlev et al.~(2004). This is enough to relax the
constraints so that core models without hyperons are viable (see Figure 6
of Yakovlev et al.~2004).

The method
for comparing the observed and predicted burst properties may also be
applicable to other bursters, although if the accretion rate exceeds a few
percent $\dot{m}_{\rm Edd}$ it would be necessary to adopt a more detailed
ignition model taking into account thermal inertia effects, like that of
\cite{woos03}.
We explicitly assume that the proportionality between the X-ray flux and
the accretion rate does not change with time. This is likely true for the
part of the 2002 October outburst of \srcb\ covered by our simulations,
given the 
monotonic relationship between the X-ray intensity and the characteristic
quasi-periodic oscillation frequencies \cite[e.g.][]{vs05}; but is not in
general true for other LMXBs.
It is also important to verify the
assumptions, particularly complete fuel consumption; such conditions are
likely the case for very regular bursts such as those typically observed
in GS~1826$-$24 \cite[]{gal03d}, but are almost certainly not the case
generally.
We note that the
degeneracy for the acceptable solutions between $X_0$ and
$Z_{\rm CNO}$ 
arises from the similarity between the H-fraction at ignition for the four
bursts, which in turn results from the relatively narrow range of $\dot{m}$
over which the bursts in \srcb\ are observed. If burst observations over a
wider range of $\dot{m}$ are available, we may very well be able to
break this degeneracy and measure the full set of compositional properties
of the burst source.

To our knowledge, this
is the first time that a sequence of bursts in 
the pure-He ignition regime has
been confirmed. At first inspection, the lightcurves of the bursts agree
well with the time-dependent models of \cite{woos03}, it will be
interesting in future work to compare the observed and theoretical
lightcurves in more detail.

\acknowledgments

We thank Lars Bildsten and Philip Podsiadlowski for useful discussions.
This research has made use of data obtained through the High Energy
Astrophysics Science Archive Research Center Online Service, provided by
the NASA/Goddard Space Flight Center.  This work was supported in part by
the NASA Long Term Space Astrophysics program under grant NAG 5-9184.
AC is an Alfred P.~Sloan Research Fellow, and is grateful for support from an NSERC Discovery grant, FQRNT, and CIAR.


\end{document}